\documentclass[aps,floatfix,superscriptaddress,twocolumn,showpacs]{revtex4}
\usepackage{amsmath,amssymb,eucal,graphicx}
\usepackage{float}
\begin{document}

\title{Temporal correlations of the running maximum of a Brownian trajectory}

\author{Olivier B\'{e}nichou}
\affiliation{Laboratoire de Physique Th\'{e}orique de la Mati\`{e}re Condens\'{e}e, UPMC,
CNRS UMR 7600, Sorbonne Universit\'{e}s, 4 Place Jussieu, 75252 Paris Cedex 05, France}
\author{P. L. Krapivsky}
\affiliation{Department of Physics, Boston University, Boston, Massachusetts 02215, USA}
\author{Carlos Mej\'ia-Monasterio}
\affiliation{Laboratory of Physical Properties, Technical University of Madrid, Av. Complutense s/n 28040 Madrid, Spain}
\author{Gleb Oshanin}
\affiliation{Laboratoire de Physique Th\'{e}orique de la Mati\`{e}re Condens\'{e}e, UPMC,
CNRS UMR 7600, Sorbonne Universit\'{e}s, 4 Place Jussieu, 75252 Paris Cedex 05, France}

\begin{abstract}
We study the correlations between the maxima $m$ and $M$ of a Brownian motion (BM) on the time intervals $[0,t_1]$ and $[0,t_2]$, with $t_2>t_1$. We determine exact forms of the distribution functions $P(m,M)$ and $P(G = M - m)$, and calculate the moments $\mathbb{E}\{\left(M - m\right)^k\}$ and  the cross-moments $\mathbb{E}\{m^l M^k\}$ with arbitrary integers $l$ and $k$. We show that correlations between $m$ and $M$ decay as $\sqrt{t_1/t_2}$ when $t_2/t_1 \to \infty$, revealing strong memory effects in the statistics  of the BM maxima. We also compute the Pearson correlation coefficient $\rho(m,M)$, the power spectrum of $M_t$, and we discuss a possibility of extracting the ensemble-averaged diffusion coefficient in single-trajectory experiments using a single realization of the maximum process. 

\end{abstract}


\maketitle

Brownian motion (BM) is a paradigmatic stochastic process \cite{levy,ito,peres,red} with enumerable applications in physics and chemistry \cite{ralf2,book}, biology \cite{berg}, computer science \cite{maj}, mathematical finance \cite{bou,chiche}, etc. Much effort has been invested in understanding the extreme value statistics (EVS) of BM, e.g., maximal or minimal displacements, spans, survival probabilities, persistence and various first-passage-time characteristics. Such results appear in numerous studies, see e.g. Refs.~\cite{katja,paul,schehr2,schehr3,ol1,ol2,kafri,carlos,ol3,ol4,sid2,alberto,david,sid,ralf,schehr,satya,mounaix,Bmaxima,2maxima,al1,al2} emphasizing the relevance of the EVS in diverse physical phenomena.

\begin{figure}[t]
\includegraphics[width=0.44\textwidth]{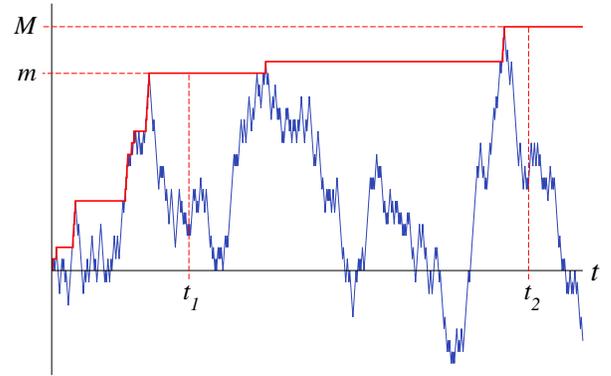}
\caption{(color online) A realization of a BM (blue) and the corresponding maximum process $M_t$ (red). $M$ and $m$ are the maxima of BM on  $[0,t_2]$ and $[0,t_1]$, respectively.}
\label{Fig1}
\end{figure}

To the best of our knowledge, nothing is known about temporal correlations of different extremes of BM, although it is interesting to probe how a maximum (minimum) achieved on a certain time interval is correlated to an extremum achieved on a longer time interval, how the span is correlated at different time moments,  how the first and the subsequent passage times depend on each other, etc. Here we address these conceptually important questions focussing on the running maximum $M_t = {\rm max}_{0 \leq s \leq t} B_s$ of a one-dimensional 
BM trajectory $B_s$ with $B_0 = 0$. We shortly write
\begin{equation*}
m = {\rm max}_{0 \leq s \leq t_1} B_s, \quad M = {\rm max}_{0 \leq s \leq t_2} B_s,\quad t_1<t_2
\end{equation*}
for the maxima achieved on the time interval $[0,t_1]$ and a longer time interval $[0,t_2]$  (see Fig.~\ref{Fig1}).  Our main goals are to determine $P(m,M)$, the joint probability distribution function (pdf) of the maxima, and the pdf $P(G)$ of the gap $G = M - m$ between the maxima. These pdfs allow us to calculate the cross-moments $\mathbb{E}\{m^l  M^k\}$, with arbitrary integer $l$ and $k$, and the moments $\mathbb{E}\{(M-m)^k\}$ of arbitrary order $k > 0$. We will show that $m$ and $M$ decouple on much larger time scales than the positions of the BM, 
revealing strong memory effects in the EVS of the BM. Using our results we extract the Pearson correlation coefficient and determine the power spectrum of $M_t$. Finally, we discuss the possibility of extracting the ensemble-averaged diffusion coefficient $D$ in single-trajectory experiments using a single realization of $M_t$. 
  
We start by summarizing a few key properties of $M_t$ which we shall need. 
Denote by $Q_t(M)$ the pdf of the maximum $M$ of BM 
on $[0,t]$. This pdf is the one-sided Gaussian distribution (see, e.g., \cite{levy,ito,peres}) 
\begin{equation}
\label{Q}
Q_t(M) = \dfrac{1}{\sqrt{\pi D t}} \exp\left(- \dfrac{M^2}{4 D t}\right) \,.
\end{equation}
Using \eqref{Q} one can express the moments $\mathbb{E}\left\{M_t^k\right\}$, with arbitrary $k>-1$, through the gamma function:
\begin{equation}
\label{maxk}
\dfrac{\mathbb{E}\left\{M_t^k\right\}}{\left(4 D t\right)^{k/2}} = \Gamma\left(\dfrac{k+1}{2}\right)/\sqrt{\pi} \,,
\end{equation}
Next, let
$\Pi_t(M,x)$ be the pdf that the BM is at $x$ at time $t$ and it has achieved the maximum $M$ during the time interval $[0,t]$. This pdf  reads (see, e.g., \cite{levy,ito,peres}) 
\begin{eqnarray}
\label{maxtail}
\Pi_t(M,x) = \dfrac{2 M - x }{2 \sqrt{\pi D^3 t^3}} \exp\!\left[- \dfrac{\left(2 M - x\right)^2}{4 D t}\right] \,.
\end{eqnarray}

\begin{figure}[t]
\includegraphics[width=0.54\textwidth]{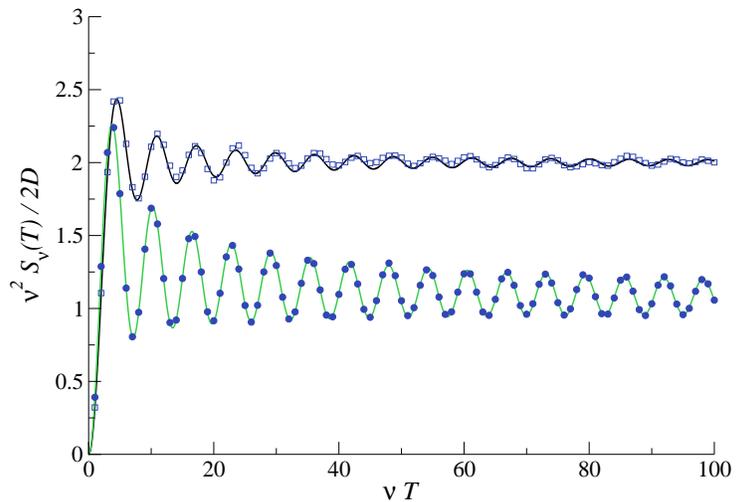}
\caption{(color online) Comparison of the power spectra of the maximum process [solid line, \eqref{ps}]
and $S^{(BM)}_{\nu}(T)$ for the BM [dashed line]. Symbols denote the results of MC simulations.
}
\label{Fig2}
\end{figure}

To determine the joint pdf $P(m,M)$ we will need an auxiliary probability $S_t(m)$ that the BM will not reach a fixed level $m > 0$ within the time interval $[0,t]$. This probability is well-known \cite{levy,ito,peres,red} 
\begin{equation}
\label{surv}
S_t(m) = {\rm erf}\!\left(\dfrac{m}{\sqrt{4D t}}\right).
\end{equation}
Here ${\rm erf}(\cdot)$ is the error function. The joint pdf $P(m,M)$ can be expressed as the sum of two contributions. The first is due to trajectories $B_s$ which reach a maximal value $m$ for $s \in [0,t_1]$,  appear at some position $x \leq m$ at $s = t_1$, and then reach a maximal value $M > m$ for $s \in [t_1,t_2]$ (see Fig.~\ref{Fig2}); the second is due to trajectories $B_s$ which reach a maximal value $m$ for $s \in [0,t_1]$,  appear at some position $x \leq m$ at $s = t_1$, and in the following time interval $s \in [t_1,t_2]$ do not reach $m$ again, so that $M=m$.  We thus formally represent $P(m,M)$ as
\begin{align}
\label{pp}
&P(m,M) = \int_{-\infty}^m dx \, \Pi_{t_1}\left(m, x\right) \, Q_{t_2 - t_1}\left(M - x\right) \nonumber\\
& + \delta\left(M - m\right) \int^m_{-\infty} dx \, \Pi_{t_1}\left(m, x\right) \, S_{t_2-t_1}\left(m-x\right)  \,.
\end{align}
Using the definitions in \eqref{Q}, \eqref{maxtail} and  \eqref{surv}, and performing the integrals in \eqref{pp}, we find the following exact result:
\begin{align}
\label{pmM}
&P\left(m,M\right) = \dfrac{\left(2 m - M\right)}{2 \sqrt{\pi D^3 t_2^3}} \exp\left(- \dfrac{\left(M - 2 m\right)^2}{4 D t_2}\right)  \times \, \nonumber\\
&{\rm erfc}\left(\sqrt{\dfrac{t_2 - t_1}{D t_1 t_2}} \, \dfrac{m}{2} + \sqrt{\dfrac{t_1}{D t_2 \left(t_2 - t_1\right)}} \, \dfrac{\left(M -  m\right)}{2} \right) \nonumber\\
&+ \dfrac{1}{\pi D t_2} \sqrt{\dfrac{t_2 - t_1}{t_1}} \exp\left(- \dfrac{m^2}{4 D t_1} - \dfrac{\left(M - m\right)^2}{4 D \left(t_2 - t_1\right)}\right) \nonumber\\
&+\dfrac{\delta\left(M - m\right)}{\sqrt{\pi D t_2}} \exp\left(- \dfrac{m^2}{4 D t_2}\right) 
 {\rm erfc}\left(\sqrt{\dfrac{t_2 - t_1}{D t_1 t_2}} \, \dfrac{m}{2}\right)  \,,
\end{align}
where ${\rm erfc}(\cdot)$ is the complementary error function. 

Equation \eqref{pmM} is our central result which allows for a direct calculation of all other properties of interest. 
For instance, using \eqref{pmM} we determine $P(G)$, the probability density that  $M-m=G \geq 0$:
\begin{align}
\label{gap}
P(G) 
&= \dfrac{2}{\pi} \arcsin\left(\sqrt{\dfrac{t_1}{t_2}}\right) \delta\left(G\right) \nonumber\\
&+ \dfrac{e^{-G^2/(4Dt_2)}}{\sqrt{\pi D t_2}}\,
{\rm erfc}\left(\sqrt{\dfrac{t_1}{D t_2 \left(t_2 - t_1\right)}} \, \dfrac{G}{2}\right) 
\end{align} 
The pdf of the gap between the first and the second ordered maxima of a BM (a different quantity from the one  we consider) has been analyzed in Ref.~\cite{mounaix}.

Next, we determine the cross-moments of the maxima $m$ and $M$ by simply integrating $P(m,M)$ in \eqref{pmM}:
\begin{align}
\label{crossmoments}
&\dfrac{\mathbb{E}\left\{m^{l} M^{k}\right\}}{(4 D t_2)^{(l+k)/2}} = \dfrac{z^{(k+l)/2}}{\pi} \Bigg[ \sum_{n=0}^k \binom{k}{n} \Gamma\left(\gamma - \dfrac{n}{2}\right)  \times \nonumber\\ 
&\Gamma\left(\dfrac{n+1}{2}\right)
\left(\dfrac{1-z}{z}\right)^{n/2}
-  \dfrac{k \, 2^{k} \, \Gamma\left(\gamma + \dfrac{1}{2}\right)}{4 \gamma}  \left(1 - z\right)^{\gamma} \times \nonumber\\
& \sqrt{z} \, \sum_{n=0}^{k-1}  \binom{k-1}{n} 
  \sum_{p=0}^l \binom{l}{p} \, \dfrac{(z-1/2)^n \, z^p \, Q_{n,p} }{\left(1 - z\right)^{2 \mu}\left(\gamma - \mu\right)}\Bigg] \,,
\end{align}
with
\begin{align}
&Q_{n,p} = \dfrac{\gamma}{\mu} \,_2F_1\left(\gamma+\dfrac{1}{2}, \mu; \mu+1; \dfrac{z}{z-1}\right) \nonumber\\
&+  \dfrac{(-1)^{n+p} \gamma}{\mu} \left(\dfrac{1-z}{z}\right)^{2 \mu} \,_2F_1\left(\gamma+\dfrac{1}{2}, \mu; \mu+1, \dfrac{z - 1}{z}\right) \nonumber\\
&- (-1)^{n+p}  \left(\dfrac{1-z}{z}\right)^{2 \mu} \,_2F_1\left(\gamma+\dfrac{1}{2}, \gamma; \gamma+1; \dfrac{z - 1}{z}\right) \nonumber\\
& - \,_2F_1\left(\gamma+\dfrac{1}{2}, \gamma; \gamma+1; \dfrac{z}{z - 1}\right) \,,
\end{align}
where $_2F_1$ denotes the hypergeometric function and
\begin{equation*}
\gamma = \frac{k + l + 1}{2}\,, \quad \mu =\frac{n + p + 1}{2}\,, \quad z = \frac{t_1}{t_2}
\end{equation*}
The first few cross-moments read
\begin{equation}
\label{m-M}
\begin{split}
\dfrac{\mathbb{E}\left\{m \, M\right\}}{2 \, D \, t_2} =\,& \dfrac{z}{2} + \dfrac{\sqrt{z \left(1 - z\right)}}{\pi} + \dfrac{1}{\pi} \arcsin\left(\sqrt{z}\right)\\
\dfrac{\mathbb{E}\left\{m \, M^{2}\right\}}{(4 \, D \, t_2)^{3/2}} = \,& \dfrac{z^{3/2} + 3 \, \sqrt{z} + 2 \, - 2 \, \left(1 - z\right)^{3/2}}{6 \, \sqrt{\pi}}\\
\dfrac{\mathbb{E}\left\{m^2 \, M\right\}}{(4 \, D \, t_2)^{3/2}} = \,& \dfrac{2 \, z^{3/2} + 1 - \left(1 - z\right)^{3/2}}{3 \, \sqrt{\pi}} \\
\dfrac{\mathbb{E}\left\{m^2 \, M^2\right\}}{8 (D \,  t_2)^2} = \,&  \dfrac{z \left(1 + z\right)}{2} + \dfrac{\left(2 z - 1\right) \sqrt{z \left(1 - z\right)}}{\pi} \\
&+ \dfrac{1}{\pi} \arcsin\left(\sqrt{z}\right)
\end{split}
\end{equation}
To highlight the decay of correlations between 
$m$ and $M$ when $t_2 \to \infty$ and $t_1$ is kept fixed, 
we formally rewrite (taking advantage of \eqref{maxk}) the first expression in \eqref{m-M} as
\begin{equation}
\frac{\mathbb{E}\left\{m \, M\right\}}{\mathbb{E}\left\{m\right\} \mathbb{E}\left\{M\right\}} 
= 1 + \dfrac{\pi}{4} \sqrt{z} + O(z) \,,
\end{equation}
implying that correlations decouple slowly, as $\sqrt{t_1/t_2}$.

From \eqref{gap} we find that the moments of the gap 
\begin{align}
\label{Gk}
&\dfrac{\mathbb{E}\left\{G^k\right\}}{\left(4 D t_2 \right)^{k/2}} \equiv \dfrac{\mathbb{E}\left\{\left(M - m\right)^k\right\}}{\left(4 D t_2 \right)^{k/2}} =  \dfrac{\Gamma\left(\dfrac{k+1}{2}\right)}{\sqrt{\pi}} 
- \nonumber\\
&- \dfrac{k \, \Gamma\left(\dfrac{k}{2}\right)  \sqrt{z} 
 \left(1 - z\right)^{(k+1)/2}}{\pi}
 \,_2F_1\left(1,\dfrac{k}{2}+1;\dfrac{3}{2};z\right)
\end{align}
for arbitrary $k>-1$. 
For example, for $k=2$ we have
\begin{eqnarray}
\label{13}
&&\mathbb{E}\left\{\left(M-m\right)^2\right\} = \dfrac{4 D t_2}{\pi} \left(\arccos\left(\sqrt{z}\right) - \sqrt{z \left(1 - z\right)}\right)  \nonumber\\
&&= \mathbb{E}\left\{M^2\right\} \left(1 -\dfrac{4}{\pi} \sqrt{z} + O\left(z^{3/2}\right)\right) 
\end{eqnarray}
which implies that the memory of $m$ fades as $\sqrt{t_1/t_2}$. The 
correlations between positions of the BM itself, $\mathbb{E}\{(B_{t_2} - B_{t_1})^2\} = \mathbb{E}\{B_{t_2}^2\} (1 - t_1/t_2)$, decay much faster.
Since $_2F_1(a,b;c;z) \to 1$ when $z\to 0$, Eq.~\eqref{Gk} yields  
$\frac{\mathbb{E}\{G^k\}}{\mathbb{E}\{M^k\}}=1 + O\big(\sqrt{t_1/t_2}\big)$ for any $k > - 1$ and $t_1/t_2\ll 1$.

Finally, we consider several direct applications of our exact results: a) First, we calculate the Pearson's coefficient 
$\rho = {\rm Cov}(m,M)/\sqrt{{\rm Var}(m) {\rm Var}(M)}$ which is a measure of the \textit{linear} correlation between $m$ and $M$:
\begin{align}
\label{p}
&\rho= \dfrac{\left(\dfrac{\pi}{2} \sqrt{z} - 2 + \sqrt{1 - z} + \dfrac{\arcsin\left(\sqrt{z}\right)}{\sqrt{z}}\right)}{\pi-2}  \,.
\end{align}
We observe that $\rho$ is a monotonically increasing function of $z$ and that $\rho \geq \rho_{BM}\left(B_{t_1},B_{t_2}\right) \equiv \sqrt{z}$,  where $\rho_{BM}$ is the Pearson coefficient for the BM, which again implies that  $M_t$ is more strongly correlated than the BM itself. 
 
 b) Further, for the power spectrum $S_{\nu}(T)$ of  $M_t$ we get 
 \begin{align}
 \label{ps}
& S_{\nu}(T) = \dfrac{1}{T} \, \mathbb{E}\left\{\left| \int^T_0 e^{i \nu t} \, M_t \, dt \right|^2\right\} \nonumber\\
&= \dfrac{2 D}{\nu^2} \left(1 - \dfrac{\sin\left(\nu T\right)}{\nu T} + 2 \sin\left(\dfrac{\nu T}{2}\right) J_1\left(\dfrac{\nu T}{2}\right) \right) \,,
 \end{align}
where $J_1(\cdot)$ is the Bessel function. This result (valid for any $\nu$ and $T$) can be compared with the power spectrum of the BM: $S^{(BM)}_{\nu}(T) \equiv 4 D (1 - \sin(\nu T)/\nu T)/\nu^2$ (see Fig.~\ref{Fig2}). Despite strong correlations and an intermittent character of the maximum process $M_t$,  its limiting power spectrum $S_{\nu}= \lim_{T \to \infty} S_{\nu}(T) = 2 D/\nu^2$ exhibits the same $\nu^{-2}$ decay as the BM, but the amplitude is two times smaller. This limit, however, is approached as $1/\sqrt{T}$ as compared to the $1/T$ relaxation taking place for the BM. Indeed, for $M_t$ we observe much stronger oscillations than for the BM (see Fig.~\ref{Fig2}). 

c) Lastly, we inquire about a possibility of extracting the ensemble-averaged diffusion coefficient $D$
from a single realization of the maximum process $M_t$. Recently, much effort has been invested 
in understanding how to do it
using $B_s$ itself, see e.g. \cite{saxton,berglund,greb,greb2,andr,boyer1,boyer2}. In particular, it was realized that a time-averaged functional of the form
\begin{equation}
\label{msd}
D_{msd} = \dfrac{1}{2 \tau \left(T - \tau\right)} \int^{T - \tau}_0 dt \left(B_{t+\tau} - B_{t}\right)^2 \,,
\end{equation}
where $\tau > 0$ is the time lag and $T$ the total observation time, is an efficient estimator of $D$. The point is that for the BM the variance ${\rm Var}(D_{msd})$ of the estimator \eqref{msd} vanishes with the observation time 
as $1/T$ (see e.g. \cite{greb}), which means that for any realization of $B_t$ the estimator converges to $D$ with probability $1$ as $T \to \infty$. 

On the other hand, if the BM takes place in bounded micro-domains, i.e., in cells, the limit $T \to \infty$ can not be taken safely since $B_t$ will start to feel the confinement at a certain moment and $D_{msd}$ will probe the finite-size rather than $D$. It means that the observation has to be interrupted at some $T$ when the variance of $D_{msd}$ is still finite. In this regard, it may be useful to have other tools to deduce $D$ which will work reliably at short $T$.  

Here we present an example of the estimator of $D$ which uses $M_t$ instead of $B_t$ itself, and has a variance which is \textit{independent} of $T$ and can be made arbitrarily small (e.g., smaller than experimental blur) by an appropriate tuning of some control parameter.   We note also that using $M_t$ instead of $B_s$ 
has a number of advantages: a) such an approach requires less data---keeping track of $B_t$ creates
a set of size $\sim T$, while in the case of $M_t$ one has to record only the 
events when $M_t$ changes its value, which, on average, happens only $\sqrt{T}$ times \cite{satya}; 
b) one may expect \cite{tej} that the estimators of $D$ based on a single realization of $M_t$ are less ``noisy", than those based on $B_t$, because $M_t$ already filters a great deal of fluctuations of $B_t$ (see Fig.~\ref{Fig1}).

\begin{figure}[t]
\includegraphics[width=0.46\textwidth]{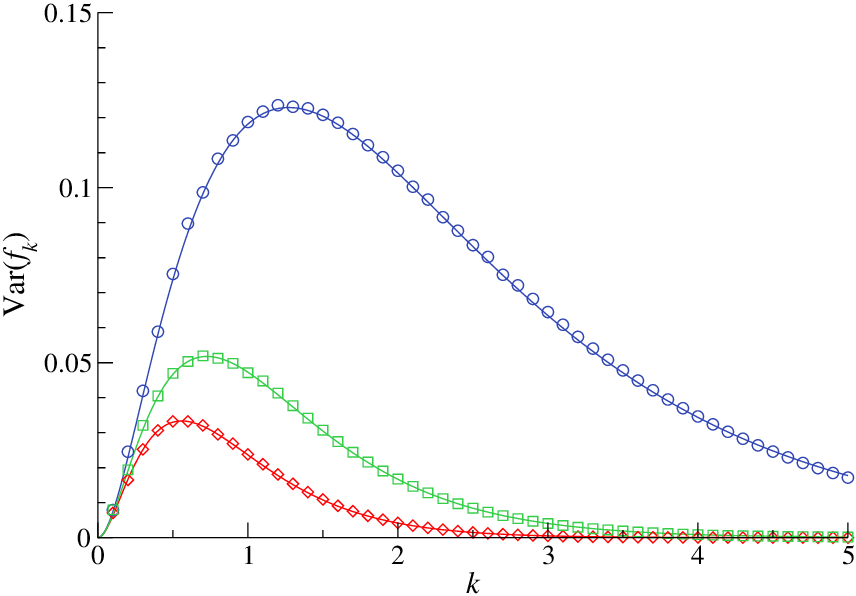}
\caption{(color online) The variance of $f_k$ as a function of  $k$ for $D=0.25$, $D=0.01$ and $D=0.005$ (from top to bottom).
Solid lines are Eq.~\eqref{variancefk} with $k$ considered as continuous variable, while the symbols are the results of the MC simulations.}
\label{FIG33}
\end{figure}

Let $B_s$ be a projection of an experimentally tracked $d$-dimensional Brownian trajectory ${\bf B}_s$ on one of the axes, and denote by $M_t$ the running maximum of this projection $B_s$.  Suppose we want to fit a random curve $M_t^k$, where $k$ is a positive number,  by some deterministic curve using the least-squares approximation.  A natural choice of the deterministic curve is provided by Eq.~\eqref{maxk} in which we replace $D$ by an estimated ``diffusion coefficient'' $D_{\it es}$. We construct  then a functional of squared residuals:
\begin{equation}
F=\int^T_0 \left(M_t^k - \Gamma\left(\dfrac{k+1}{2}\right) \dfrac{\left(4 D_{\it es} t\right)^{k/2}}{\sqrt{\pi}}\right)^2 \, dt 
\end{equation}   
Regarding $D_{\it es}$ as an optimization parameter, we minimize $F$ and find the minimum
\begin{equation}
\label{fk}
f_k \equiv D_{\it es}^{k/2} =  \dfrac{\sqrt{\pi} \left(\dfrac{k}{2}+1\right)}{2^{k} \Gamma\left(\dfrac{k+1}{2}\right) T^{k/2+1}}\, \int^T_0 M^k_t dt 
\end{equation}
providing us with a $k$-parametrized family of estimators minimizing an error in the least-squares fitting of $M^k_t$ of a given realization of $M_t$.  While the ensemble-averaged value is 
$\mathbb{E}\left\{f_k\right\} \equiv D^{k/2}$, $f_k$ fluctuates around this value giving an estimate 
$D_{\it es}^{k/2}$ of the actual value $D^{k/2}$.  To quantify the fluctuations of $f_k$ we use Eq.~\eqref{crossmoments} to
compute \cite{integer} the variance of $f_k$
\begin{equation}
\label{variancefk}
\dfrac{{\rm Var}\left(f_k\right)}{D^{k}} =  \dfrac{\sqrt{\pi} \, (k+2) \, (3 k +2 ) \, \Gamma\left(k + 1/2\right)}
{(4\Gamma[(k+3)/2])^2} - 1 \,
\end{equation}

\begin{figure}[t]
\includegraphics[width=0.45\textwidth]{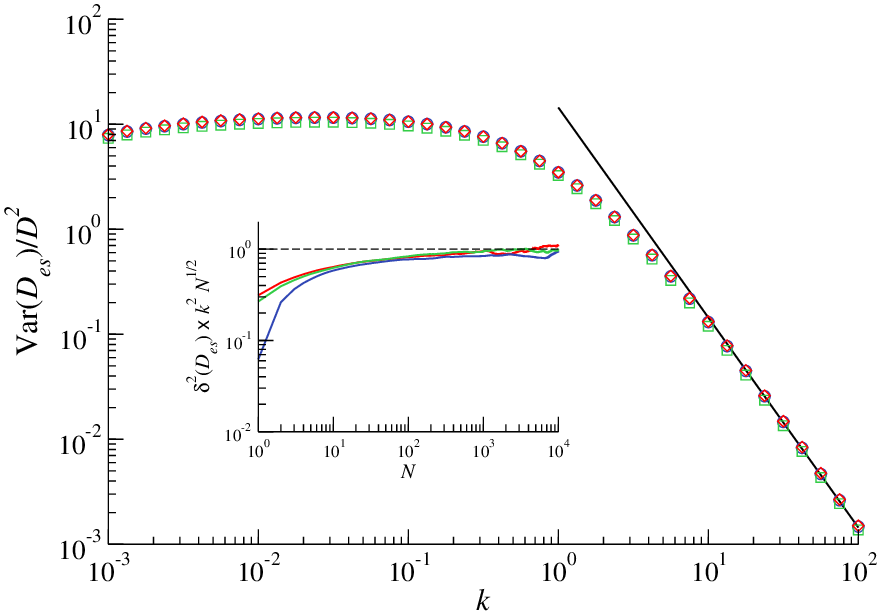}
\caption{(color online) ${\rm Var}(D_{es} = f_k^{2/k})/D^2$ 
using non-linearly transformed estimator in Eq. \eqref{fk} as a function of $k$ for $D = 0.25$, $D= 0.01$ and $D=0.005$ (from top to bottom). The solid line is $1/k^2$.  The inset shows the dependence of the 
deviation $\delta^2(D_{es})$ (see the text) on the number $N$ of recorded points of a discretised trajectory.
Different colors correspond to $k=10$ (blue), $k=50$ (green) and $k=10^2$ (red).}
\label{FIG3333}
\end{figure}

Numerical simulations indicate the validity of \eqref{variancefk} for non-negative, not necessarily \textit{integer}, values of $k$ (see Fig.~\ref{FIG33}). Inspecting Eq. \eqref{variancefk} we observe that ${\rm Var}\left(f_k\right)$ is a non-monotonic (for $D < 1/2$) function of $k$ which vanishes when $k \to 0$ or $k \to \infty$, suggesting that we have to take either very small or very big values of $k$ in order to minimize the error of the estimator in Eq.~\eqref{fk}.  We haven't been able to determine the distribution $P(f_k)$, so we resorted to numerical analysis to get the variance of the non-linearly transformed variable $D_{es} = f_k^{2/k}$. The results of our MC simulations (Fig.~\ref{FIG3333}) show that ${\rm Var}(D_{es})$ is a non-monotonic function of $k$ and it is indeed advantageous to use big values of $k$ for which this variance can be made arbitrarily small. 

\begin{figure}[t]
\includegraphics[width=0.47\textwidth]{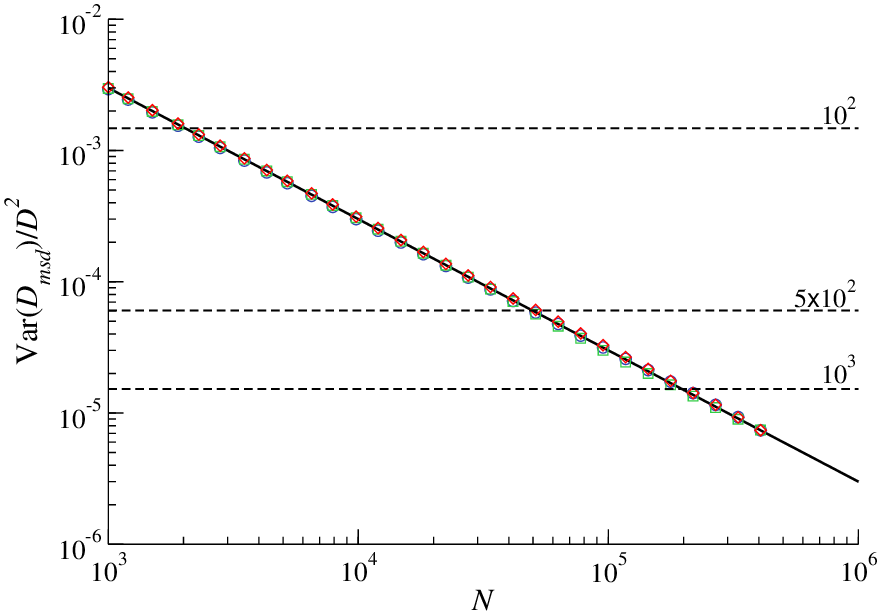}
\caption{(color online) The variance of the discretized 
time-averaged functional in Eq.~\eqref{msd}, divided by $D^2$, versus 
the number $N$ of recorded positions of the trajectory $B_t$.  
Solid line is the theoretical result, ${\rm Var(D_{msd})/D^2 \sim 3/N}$, 
while the symbols (the same as in Figs.\ref{FIG33} and \ref{FIG3333}) are results of MC simulations for $D=0.25$, $D=0.01$ and $D=0.005$. The horizontal dashed lines give ${\rm Var}(D_{es})/D^2$ for $ k=10^2$, $5 \times 10^2$ and $10^3$.}
\label{FIG333}
\end{figure}

The variance ${\rm Var}(D_{es})$ is independent of time, yet in practice one records the trajectory $B_t$ at discrete time moments and in this case ${\rm Var}_N(D_{es})$ starts to depend on the number $N$ of recorded points, attaining the limiting value ${\rm Var}(D_{es})$ when $N \to \infty$.  In the inset to Fig.~\ref{FIG33} we plot the results of the MC simulations for the deviation $\delta^2(D_{es}) = ({\rm Var}(D_{es}) - {\rm Var}_{N}(D_{es}))/D^2$ as a function of $N$ for several values of the control parameter $k$.  We observe that the curves corresponding to different values of $k$ collapse when we plot $k^2 \sqrt{N} \delta^2(D_{es})$  implying that $\delta^2(D_{es}) \sim c/(k^2 \sqrt{N})$, where $c$ is a constant of order of unity. Thus the error stemming out of a finite $N$ can be made arbitrarily small by choosing a sufficiently large value of $k$.  

 Lastly, in Fig.~\ref{FIG333} we compare the variance of the commonly used estimator in Eq.\eqref{msd}  against the variance of the estimator $D_{es}$, based on $M_t$. We observe that at short times the latter is much smaller which supports our guess that the ensemble-averaged diffusion coefficient $D$ can be reliably deduced from estimators based on $M_t$. Seeking other estimators based on extremal properties of $B_t$ which possess an ergodic property suggests an interesting new field of research.

The research of O.B. was supported by ERC grant FPTOpt-277998.
PLK is grateful to the IPhT CEA-Saclay for hospitality and excellent working conditions.
C.M-M. and G.O. acknowledge a partial support from the Office of Naval Research Global Grant N62909-15-1-C076 and wish to thank the Institute for Mathematical Sciences of  the National University of Singapore, where some preliminary work  was done, for warm hospitality and a financial support. C.M-M also acknowledges the support from the Spanish MICINN grants MTM2012-39101-C02-01 and MTM2015-63914-P.


\begin{thebibliography}{99}
\bibitem{integer}
  The usage of Eq.~\eqref{crossmoments} limits our analysis to non-negative integer values of $k$.
\end{thebibliography}
\end{document}